\documentclass[manuscript,11pt]{aastex}
\usepackage{epsfig}
\usepackage{soul}
\usepackage{color}

\def\fexxvi{Fe\,{\sc xxvi}}

\def\cm{\ifmmode {\rm cm}^{-1} \else cm$^{-1}$ \fi}
\def\s{\ifmmode {\rm s}^{-1} \else s$^{-1}$ \fi}
\def\cc{\ifmmode {\rm cm}^{-3} \else cm$^{-3}$ \fi}
\def\cs{\ifmmode {\rm cm}^{-2} \else cm$^{-2}$ \fi}
\def\g{\ifmmode \gamma \else $\gamma$\fi}
\def\G{\ifmmode \Gamma \else $\Gamma$\fi}
\def\Gs{\ifmmode \Gamma~ \else $\Gamma~$\fi}

\def\gc{\ifmmode \gamma_{\rm c} \else $\gamma_{\rm c}$ \fi}
\def\sw{Schwarzschild~}
\def\gsim{\mathrel{\raise.5ex\hbox{$>$}\mkern-14mu
             \lower0.6ex\hbox{$\sim$}}}
\def\lsim{\mathrel{\raise.3ex\hbox{$<$}\mkern-14mu
             \lower0.6ex\hbox{$\sim$}}}
\def\simless{\mathbin{\lower 3pt\hbox
     {$\rlap{\raise 5pt\hbox{$\char'074$}}\mathchar"7218$}}}   
\def\simmore{\mathbin{\lower 3pt\hbox
     {$\rlap{\raise 5pt\hbox{$\char'076$}}\mathchar"7218$}}}   
\def\Msun{M_\odot}                                
\def\4u{4U 1728--34}
\def\deg{^\circ}

\def\obj{PDS~456}

\lefthead{et al.} \righthead{\4u}

\shorttitle{Magnetic UFOs in PDS~456}

\shortauthors{Fukumura et al.}

\begin{document}

\title{Variable Nature of Magnetically-Driven Ultra-Fast Outflows}

\date{\today}

\author{\textsc{Keigo Fukumura}\altaffilmark{1},
\textsc{Demosthenes Kazanas}\altaffilmark{2}, \textsc{Chris Shrader}\altaffilmark{2,3},
\textsc{Ehud Behar}\altaffilmark{4}, \textsc{Francesco Tombesi}\altaffilmark{2,5,6}
\textsc{and} \textsc{Ioannis Contopoulos}\altaffilmark{7} }

\altaffiltext{1}{Department of Physics and Astronomy, James Madison University,
Harrisonburg, VA 22807; fukumukx@jmu.edu}
\altaffiltext{2}{Astrophysics Science Division, NASA/Goddard Space Flight Center,
Greenbelt, MD 20771}
\altaffiltext{3} {Catholic University of America, Washington, DC 20064}
\altaffiltext{4}{Department of Physics, Technion, Haifa 32000, Israel}
\altaffiltext{5}{Department of Astronomy and CRESST, University of Maryland, College
Park, MD20742}
\altaffiltext{6}{Department of Physics, University of Rome ``Tor
Vergata", Via della Ricerca Scientifica 1, I-00133 Rome, Italy}
\altaffiltext{7}{Research Center for Astronomy, Academy of Athens, Athens 11527,
Greece}

\begin{abstract}
\baselineskip=15pt

Among a number of active galactic nuclei (AGNs) that drive ionized outflows in X-rays, a low-redshift ($z = 0.184$) quasar, PDS~456, is long known to exhibit one of the exemplary ultra-fast outflows (UFOs).
However, the physical process of acceleration mechanism is yet to be definitively {constrained}. In this work, we model the variations of the Fe K UFO properties in \obj\ over many epochs in X-ray observations 
in the context of magnetohydrodynamic (MHD) accretion-disk winds employed in our earlier studies of similar X-ray absorbers. 
We applied the model to  the 2013/2014 {\it XMM-Newton}/{\it NuSTAR} spectra to determine the UFO's condition; namely, velocity, ionization parameter, column density and equivalent width (EW). 
%
Under some provisions on the dependence of X-ray luminosity on the accretion rate 
applicable to near-Eddington state, our photoionization calculations, coupled to a 2.5-dimensional MHD-driven wind model, can further reproduce the observed correlations of the UFO velocity and the anti-correlation of its EW with X-ray strength of \obj.
%
This work demonstrates that UFOs, even without radiative pressure, can be driven as an extreme case purely by magnetic interaction while also producing the observed spectrum and correlations. 

\end{abstract}

\keywords{accretion, accretion disks --- (galaxies:) quasars: absorption lines ---
methods: numerical --- galaxies: individual (\obj)  --- magnetohydrodynamics (MHD) }


\baselineskip=15pt

\section{Introduction}

One of the generic features seen in black hole (BH) systems such as active galactic nuclei (AGNs) and Galactic X-ray binaries (XRBs) are blue-shifted absorption features in their spectra primarily detected in the UV and the X-ray bands, with the latter also known as warm absorbers (WAs).
%
%
Within the last decade, on the other hand, a new class of outflows has drawn much attention because of their unique physical characteristics: They are ejected at near-relativistic velocities ($v/c \sim 0.1-0.7$), with nearly Compton-thick columns ($10^{23} \lsim N_H \lsim 10^{24}$ cm$^{-2}$) and systematically high ionization parameter\footnote[1]{This is defined as $\xi \equiv L_{\rm ion} / (n r^2)$ where $L_{\rm ion}$ is ionizing (X-ray) luminosity, $n$ is plasma number density at distance $r$ from the BH.} ($\log \xi \sim 4-6$). These ultra-fast outflows (UFOs), primarily observed with high-throughput CCD detectors, appear to be present not only in nearby Seyfert AGNs \citep[e.g.][]{Tombesi10,Tombesi11,Tombesi14} but also in very bright (lensed) quasars \citep[e.g.][]{Chartas03,Pounds03,Chartas09a,Dadina18} and presumably in the ultraluminous X-ray sources \citep[e.g.][]{Walton16}.
While WA/UFO signatures in the X-ray spectra are thought to be generic to accretion-powered sources, their launching mechanism is still poorly understood.
%

%

\obj\ is an archetypal nearby ($z=0.184$), radio-quiet quasar (QSO) hosting a BH of mass $M \sim 10^9 \Msun$ \citep[e.g.][]{Reeves09}, being the most luminous AGN in the local universe with a bolometric luminosity of $L_{\rm bol} \sim 10^{47}$ erg/s. It is among the best studied QSOs for its strong UFO signatures observed in the Fe K band in the past X-ray observations with {\it Suzaku} \citep[e.g.][]{Reeves09,Reeves14,Gofford14,Matzeu16}, {\it XMM-Newton}/EPIC and {\it NuSTAR} \citep[e.g.][hereafter, N15]{Reeves03,Behar10,Nardini15}. A number of  these extensive monitoring campaigns of \obj\ at different flux levels between 2001 and 2014 has provided us with insights into the variable nature of the UFOs 
since its first discovery \citep[e.g.][]{Reeves03}. Detailed spectral analyses have consistently confirmed that the highly ionized UFOs of $v/c \sim 0.2-0.3$ could be identified as the 1s-2p resonance transitions of \fexxvi\ with almost Compton-thick column of $N_H \lesssim 10^{24}$ cm$^{-2}$ and high ionization parameter of $\log \xi \sim 4-6$ \citep[e.g.][]{Reeves09}.
The distance of the observed UFOs is estimated to lie within $\sim 50 R_S$ where $R_S$ is the \sw radius
\citep[e.g.][]{Reeves09}. The
spectral variability has been studied as well suggesting that variable
partial covering and/or intrinsic continuum variation may cause the
observed short-term variability in X-ray \citep[][hereafter, M16]{Matzeu16}.
\citet{Reeves18} also found an even faster Fe K UFO ($v/c \sim 0.42$) in 2017 {\it
XMM-Newton/NuSTAR} observation.


The QSO outflows have been thought to be driven by radiation
pressure by their O/UV flux, in a manner similar to that observed in massive stars
\citep[][]{King10,Hagino15,KingPounds15,Hagino17,Nomura17}. However, most UFOs are so
highly ionized that there is little UV or soft X-ray opacity, making this
process very inefficient \citep[e.g.][]{Higginbottom14}. Alternatively, winds can be
driven by the action of global magnetic fields threading the accretion disk to provide
a plausible means of efficient acceleration as observed (e.g.
\citealt{BP82}, hereafter, BP82; \citealt{CL94}, hereafter, CL94; \citealt{KK94},
\citealt{F10}, hereafter F10; \citealt{K12}; \citealt{K15}; \citealt{Kraemer18}).

\obj\ UFOs are especially interesting because of the variation of its properties
with the X-ray variability over the past fifteen or so years. 
For example, \citet[][hereafter, M17]{Matzeu17} noted, over a decade of multi-epoch
observations,  that the detected Fe K UFO velocity is well correlated with X-ray
luminosity $L_X$ ($7-30$ keV) suggestive of a radiative-driven origin. Independently, \citet[][hereafter, Pa18]{Parker18} showed a likely
anti-correlation between the observed UFO EW  and $2-10$ keV flux. A similar
trend is also reported in the {\it XMM-Newton}/EPIC-pn and {\it
NuSTAR} spectra for the low-redshift narrow-line Seyfert 1 (NLS1), IRAS~13224-3809
(\citealt{Parker17}, hereafter P17; \citealt{Pinto18}, hereafter Pi18) possibly accreting at near-Eddington rate.

Motivated by these findings, we investigate in this Letter the dependence of the variation of the UFO properties on the X-ray flux within the MHD wind framework of our earlier studies. In \S 2, we overview the proposed wind models. In \S 3, we present our results and demonstrate that
magnetically-driven winds can account for the observed correlations. In \S 4 we
conclude with a summary and discussion.

\section{MHD-Driven Wind Model}

The magnetically-launched disk-wind model 
has been applied to account
successfully for the X-ray absorber properties of sources of very wide mass range; i.e. the galactic XRB GRO~J1655-40 \citep[][]{F17} to the Seyfert 1
NGC~3783 \citep[][]{F18} and a nearby QSO, PG~1211+143
\citep[][]{F15}, thus establishing its scale-invariance with BH mass and its broad
applicability.
{\bf A detailed model description can be found in the above references and we will briefly review the essential elements of the model below. }
%
%

\obj , because of its very high accretion rate, may present an uncertainty in
relating the ionizing luminosity $L_{\rm ion}$ (13.6 eV - 13.6 keV) to the mass-accretion rate
$\dot{m}_a$, provided by its large O/UV luminosity. We thus parameterize this as $L_{\rm ion} \equiv f_{44} 10^{44}$ erg~s$^{-1}$ such that $f_{44} \propto \dot{m}_a^s$  with $s<1$ since a part of the radiation  is likely to be trapped in the flow and advected into the BH for $\dot m_a \gsim 1$ \citep[e.g.][for supercritical accretion]{Takeuchi09}.
We introduce the fractional mass outflow rate to  mass accretion rate, $\eta_w$,
such that $\eta_w \propto \dot{m}_a$ rather than being constant, allowing the wind mass flux to vary.
The wind density is given by $n_w(r,\theta) = n_{o}^{w}
(r/R_S)^{-p} g(\theta)$ where $n_o^{w}$ denotes the wind density normalization (i.e.
wind density at its innermost launching radius on the disk surface at $r \gsim R_S$),
with the index $p = 1.2$ observationally consistent with our earlier work \citep{B09,F17,F18} .
Note that $g(\theta)$ is the (polar) angular dependence {\bf to be solved by the model (see CL94, F10)}.
Since the plasma optical depth on the disk surface is proportional to $\sigma_T R_S
n_o$ where $\sigma_T$ is the Thomson cross section, it is found that
\begin{eqnarray}
n_o \sim \frac{\tau}{\sigma_T R_S} \propto \frac{\dot{m}_a}{\sigma_T R_S} \ ,
~~ n_o^{w} \sim \frac{\eta_w \dot{m}_a}{\sigma_T R_S} \  .\label{eq:n0}
\end{eqnarray}
One can then express $\dot{m}_a$ in terms of $f_{44}$ to yield
%
$n_o^w \propto \eta_w \dot{m}_a \propto \dot{m}_a^2  \propto f_{44}^{2/s} \equiv n_{10} 10^{10} ~\textmd{cm$^{-3}$}$
%
where $n_{10}$ is to be constrained by observations.
%
%
%
Hence, the ionization parameter $\xi$ scales with $f_{44}$ as
%
$\xi   \propto f_{44}/n_o^w  \propto f_{44}^{1-2/s}$.
With this scalings the wind density increases faster than the X-ray flux for $s<2$, and
therefore, increase in $L_{\rm ion}$ will bring the wind ionization front to smaller radii,
yielding higher plasma velocities.


While the wind kinematics is fully determined by the ideal MHD
equations (e.g. BP82, \citealt{KK94}, CL94, \citealt{Kraemer18}), we consider a
possibility of truncated disk, as often discussed  \citep[e.g.][]{Nemmen14,Hogg18}, by
introducing an {\bf inner truncation launching radius}, $R_T$.
{\bf This parameter is constrained by fitting the observed UFO spectrum. }
Given the presence of the big-blue-bump \citep[generally attributed to the accretion disk; see,][]{Matzeu16}, we assume a fiducial value of
$\theta_{\rm obs}=50\deg$ for the inclination angle, as also previously suggested
\citep[e.g.][]{Reeves09,Reeves14,Hagino15}. 
{\bf
It should be kept in mind that no radiation is deliberately taken into account in this work to illustrate the pure magnetic case.
}

%

Employing the self-similar prescription in the radial direction with the Keplerian velocity profile  ($v \propto r^{-1/2}$), the poloidal field structure is
determined by numerically solving the Grad-Shafranov equation as is originally
formulated in CL94. Hence, the wind
geometry is inherently 2.5D.
The outflowing plasma is then photoionized by the radiation of
spectral shape $F_\nu$ and luminosity $L_{\rm ion}$ assumed to be a compact region much smaller than the UV-emitting
region \citep[e.g.][]{Chartas09b,Morgan12}. We adopt the input SED of \obj\ from M16
where simultaneous observations with {\it XMM-Newton}/OM and {\it NuSTAR} in 2014 are
phenomenologically parameterized in a double broken power-law form; i.e. $\Gamma = 0.7$
for O/UV - 10 eV, $\Gamma =3.3$ for soft X-ray band and $\Gamma=2.4$ beyond $0.5$ keV
(see also N15).

%
%
%
%
In response to the irradiating SED, the wind ionization is computed as described in
F10, by employing {\tt xstar} \citep[][]{KallmanBausas01} to calculate the local ionic
abundances and the photo-excitation cross section $\sigma_{\rm abs}$. 
The latter is a function of local wind velocity $v(r,\theta)$ and its radial shear $v_{\rm
shear}(r,\theta)$; the wind shear is implemented in the Voigt function $H(a,u)$ (see
F10) to effect a physically motivated local line broadening consistent with the wind
kinematics, instead of the arbitrary choice of a
{\it turbulent} velocity (unphysical in its magnitude considering the much smaller
thermal plasma velocities). A local line depth is calculated by $\tau_\nu(r,\theta)=\sigma_{\rm abs} N_{\rm ion}$ where $N_{\rm ion}$ is the local ionic column over a discretized small distance $\Delta r ~(\Delta r/r = 0.15)$ along a line-of-sight (LoS).
The observed spectrum is then a superposition of all local spectra over the entire
wind (F10). The strength of the UFO feature is measured by equivalent width (EW) as
\begin{eqnarray}
\textmd{EW}(\theta) \equiv \int_{\rm wind} \left[1-\prod_{\rm wind} e^{-\tau_{\nu}(r,\theta)} \right]d \nu \ .
\end{eqnarray}

\section{Results}

\subsection{UFOs in 2013/2014 Composite Spectrum}

We first attempt to model a composite 
{\it XMM-Newton/NuSTAR} spectrum in 2013/2014 
where these observations were close together in time (4 days) and very little variability is
noted. A detailed analysis and discussion are found in N15.

\begin{figure}[h]
\begin{center}$
\begin{array}{cc}
\includegraphics[trim=0in 0in 0in
0in,keepaspectratio=false,width=3.4in,angle=-0,clip=false]{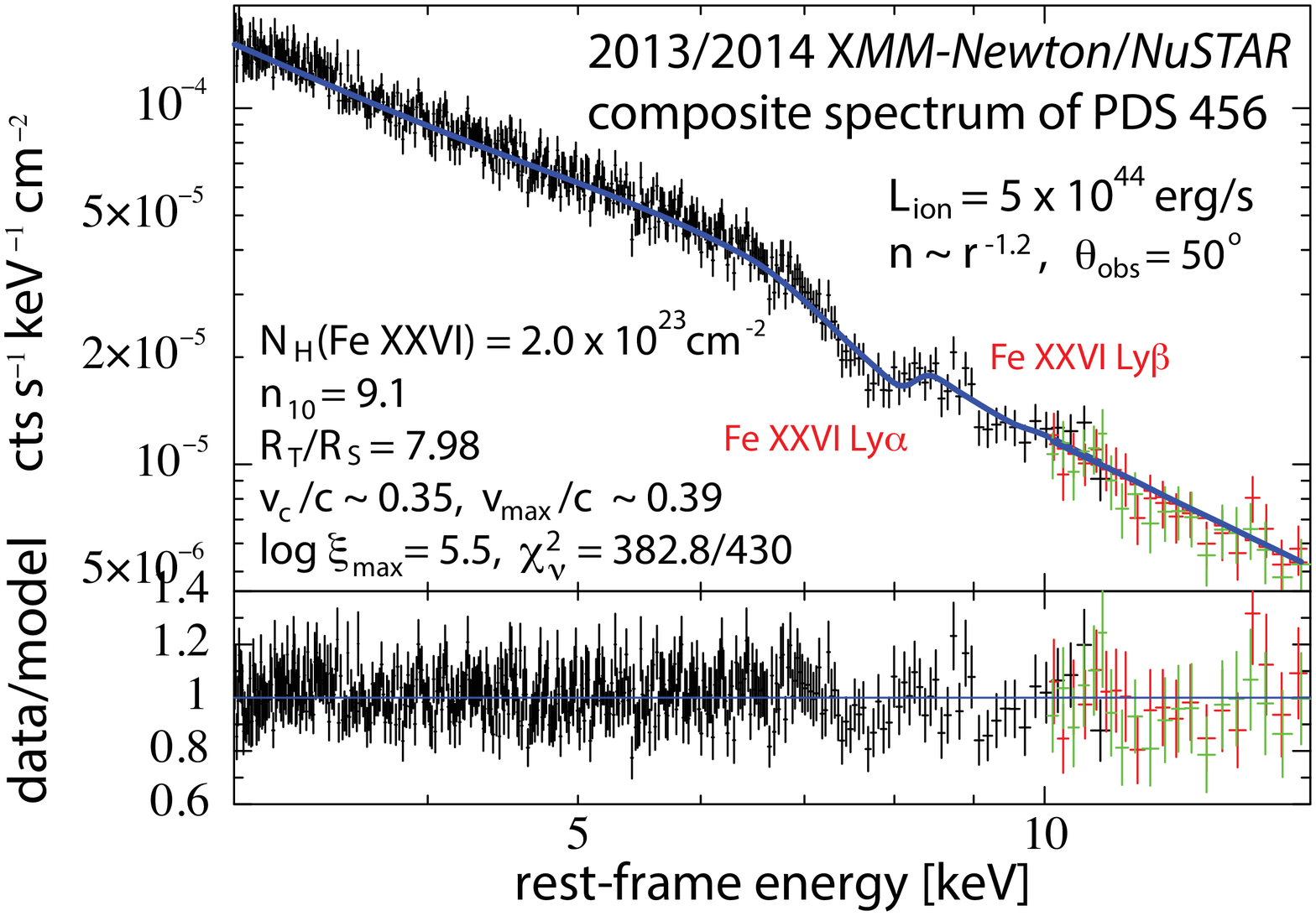}\includegraphics[trim=0in 0in 0in
0in,keepaspectratio=false,width=3.2in,angle=-0,clip=false]{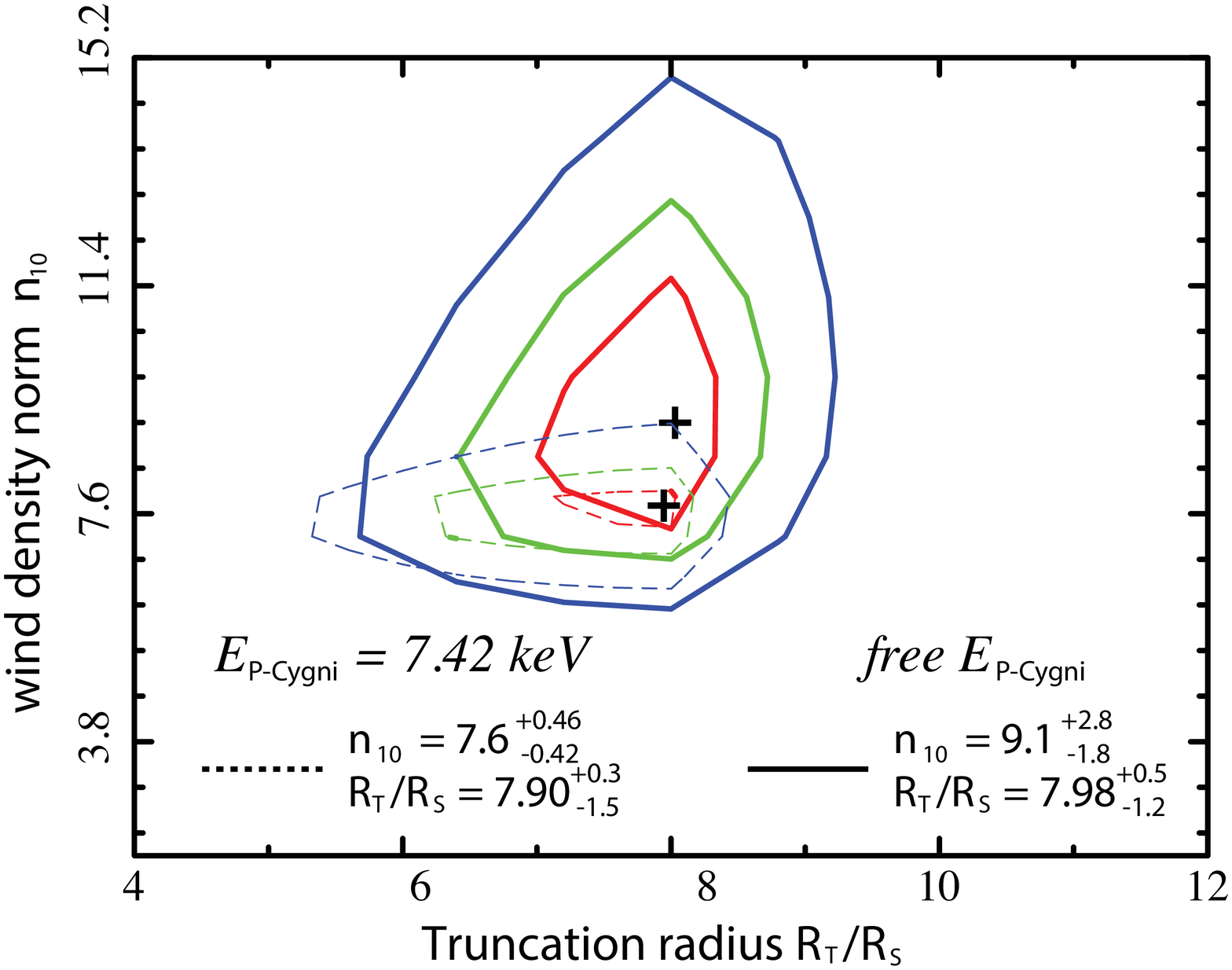}
\end{array}$
\end{center}
\caption{(a) A composite {\it XMM-Newton}/EPIC-pn/mos (black) and {\it NuSTAR} (red and green)
spectrum of \obj\ (N15) showing the observed Fe K UFOs fitted with the MHD-wind model (blue)
with a $p=1.2$ density profile assuming $\theta_{\rm obs}=50\deg$.
(b) Confidence contours showing $68\%$ (red), $90\%$ (green) and $99\%$ (blue) levels
with the bestfit solution (cross in black) with $E_{\rm P-Cygni}=7.42$ keV as derived
from N15 (dashed) and the {\it free} $E_{\rm P-Cygni}$ (solid). }
\label{fig:wind1}
\end{figure}

We analyze the spectrum of \obj\ {\bf by assuming a canonical density slope $p=1.2$,
as discussed above.}
Adopting the galactic absorption model with {\tt tbabs} where $N_H^{\rm Gal} = 2.4
\times 10^{21}$ cm$^{-2}$ \citep[][]{Kalberla05}, we initially excluded the Fe K band
($\sim 7$ keV $\lsim E \lsim$ $\sim 11$ keV in the rest-frame) to model the underlying
continuum, following the earlier analysis by N15.
As explained in \S 2, for a given X-ray luminosity of $f_{44}=5$ (i.e. $L_{\rm ion}
\sim 5 \times 10^{44}$ erg/s) in this specific epoch (\citealt{Gofford14}; N15; M17),
our primary model parameters include: (1) a wind truncation radius $R_T$ and (2) the
density normalization $n_{10}$, as shown in {\bf Table~1} where we systematically explored
the parameter space that yields the bestfit spectrum. To account for the
presence of a broad P-Cygni profile at Fe K band (see N15), a single gaussian component
{\tt zga} is (phenomenologically) added. 
With the above continuum,
the absorption signature is fitted by our MHD wind model {\tt mhdwind}.
This component significantly improves the fit with $\chi^2$/dof=$382.8/430$ with 
$R_T/R_S=7.98^{+0.5}_{-1.2}$ and $n_{10}=9.1^{+2.8}_{-1.8}$ when $E_{\rm P-Cygni}$
is freely varied, while $\chi^2$/dof=$501.1/429$ for $R_T/R_S=7.90^{+0.3}_{-1.5}$ and
$n_{10}=7.6^{+0.46}_{-0.42}$ when fixed at $E_{\rm P-Cygni}=7.42$ keV as constrained in
N15. The former bestfit yields $v_c/c \simeq 0.35$ (the centroid velocity) and $v_{\rm
max}/c \simeq 0.39$ (innermost wind velocity). 
{\bf Figure~\ref{fig:wind1}a} shows  the {\it XMM-Newton}/EPIC (black) and {\it NuSTAR}
(red/green) composite spectrum of \obj\ modeled with the MHD disk-wind 
(blue) with free $E_{\rm P-Cygni}$. Considering the implied high-velocity and
ionization parameter, we only focus on \fexxvi\ ion in this work. The Ly$\beta$ line, while modeled here, is
insignificant due to its lower oscillator strength as expected. 

A radiative transfer calculation using {\tt xstar} in thermal/ionization equilibrium
also provides us with a number of important wind quantities; i.e. the plasma
temperature $T$ and ionization parameter $\xi$ and the hydrogen-equivalent column
density $N_H$ for the UFO, as listed in {\bf Table~2}.
%
%
It is seen that the \fexxvi\ is  formed predominantly in the innermost
wind layer close to the truncated radius. The obtained bestfit solution $(n_{10}, R_T)$
is well constrained as demonstrated in the confidence contour in {\bf
Figure~\ref{fig:wind1}b}. 
{\bf We note that the bestfit result is almost independent of our choice of $p=1.2$. }
Being encouraged by the successful modeling of the Fe UFO seen in the 2013/2014 data,
we then investigate a variable nature of the reported UFO conditions in \S 3.2.


\begin{deluxetable}{l|c}
\tabletypesize{\small} \tablecaption{Primary Grid of MHD-Wind Model Parameters } \tablewidth{0pt}
\tablehead{Primary Parameter & Value}
\startdata
Truncation radius $R_T$ [in $R_S$] & $0, 2, 4, 8, 16, 32, 64, 128$ \\
Wind density normalization $n_{10}$$^{\dagger}$  & 0.01 - 40 \\
\enddata
\vspace{0.05in}
We assume $M = 10^9 \Msun$ \citep{Reeves09}, $\theta_{\rm obs}=50\deg$ and $p=1.2$.
\\
$^\dagger$ Wind density normalization in units of $10^{10}$ cm$^{-3}$.
\label{tab:tab1}
\end{deluxetable}

\subsection{The UFO Correlations}

Following the successful model fit to the 2013/2014 Fe K UFOs for \obj\ in \S 3.1, we
investigated the dependence of the UFO properties (i.e. $v$ and EW) on the ionizing
X-ray luminosity depicted by $f_{44}$, while holding everything else constant for
simplicity. It should be reminded that the radiation field plays no role in affecting
outflow kinematics in this model. On the other hand, its ionization structure is greatly influenced. As studied in M17, we consider a variable luminosity of $1 \lesssim f_{44} \lesssim 26$
with $\theta=50\deg$ assuming that the wind geometry changes very little with changing
$f_{44}$ if the underlying global magnetic field is sufficiently ``stiff" against the
change in radiation field. 

\begin{figure}[h]
\begin{center}$
\begin{array}{cc}
\includegraphics[trim=0in 0in 0in
0in,keepaspectratio=false,width=3.4in,angle=-0,clip=false]{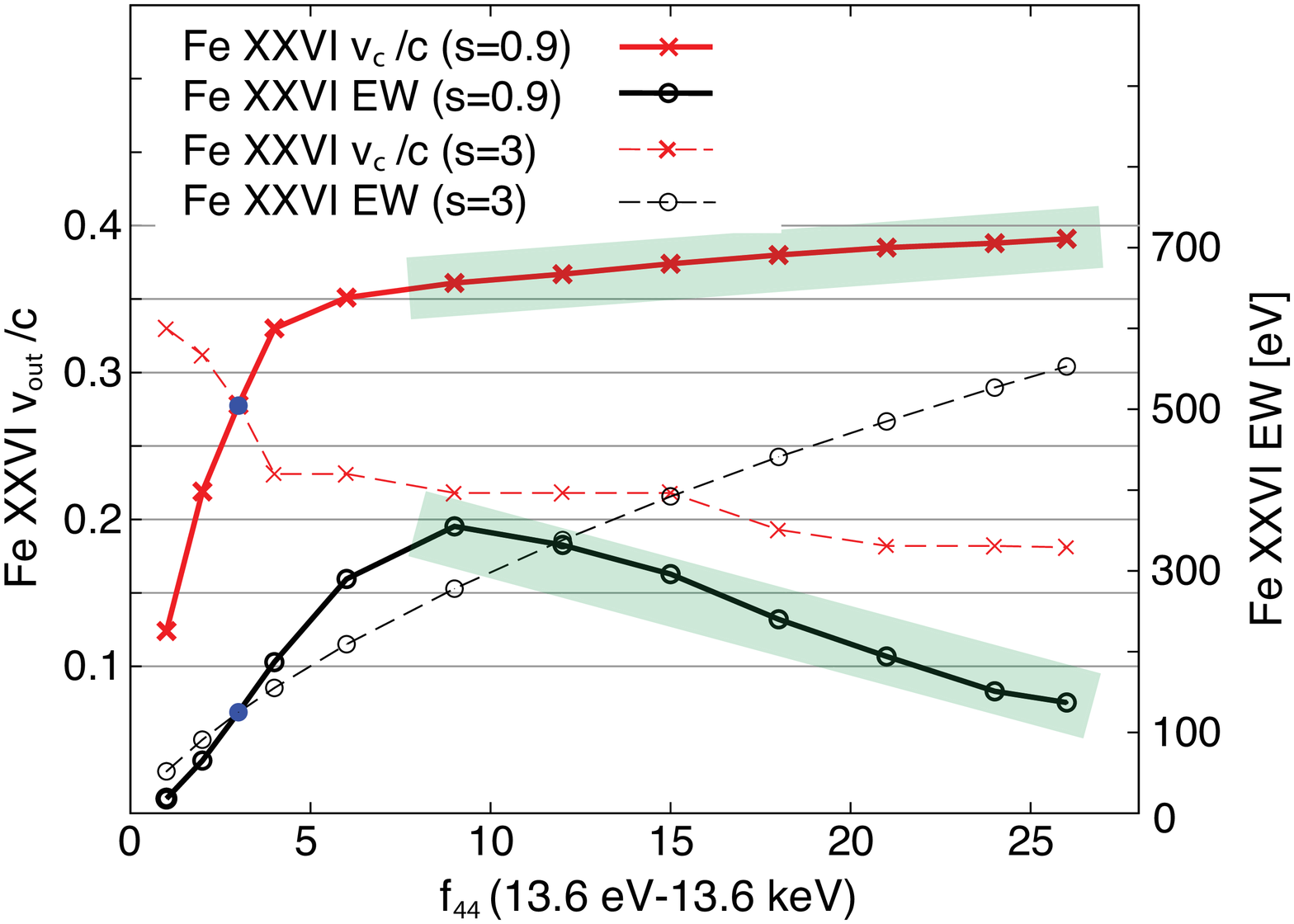}
\includegraphics[trim=0in 0in 0in
0in,keepaspectratio=false,width=3.3in,angle=-0,clip=false]{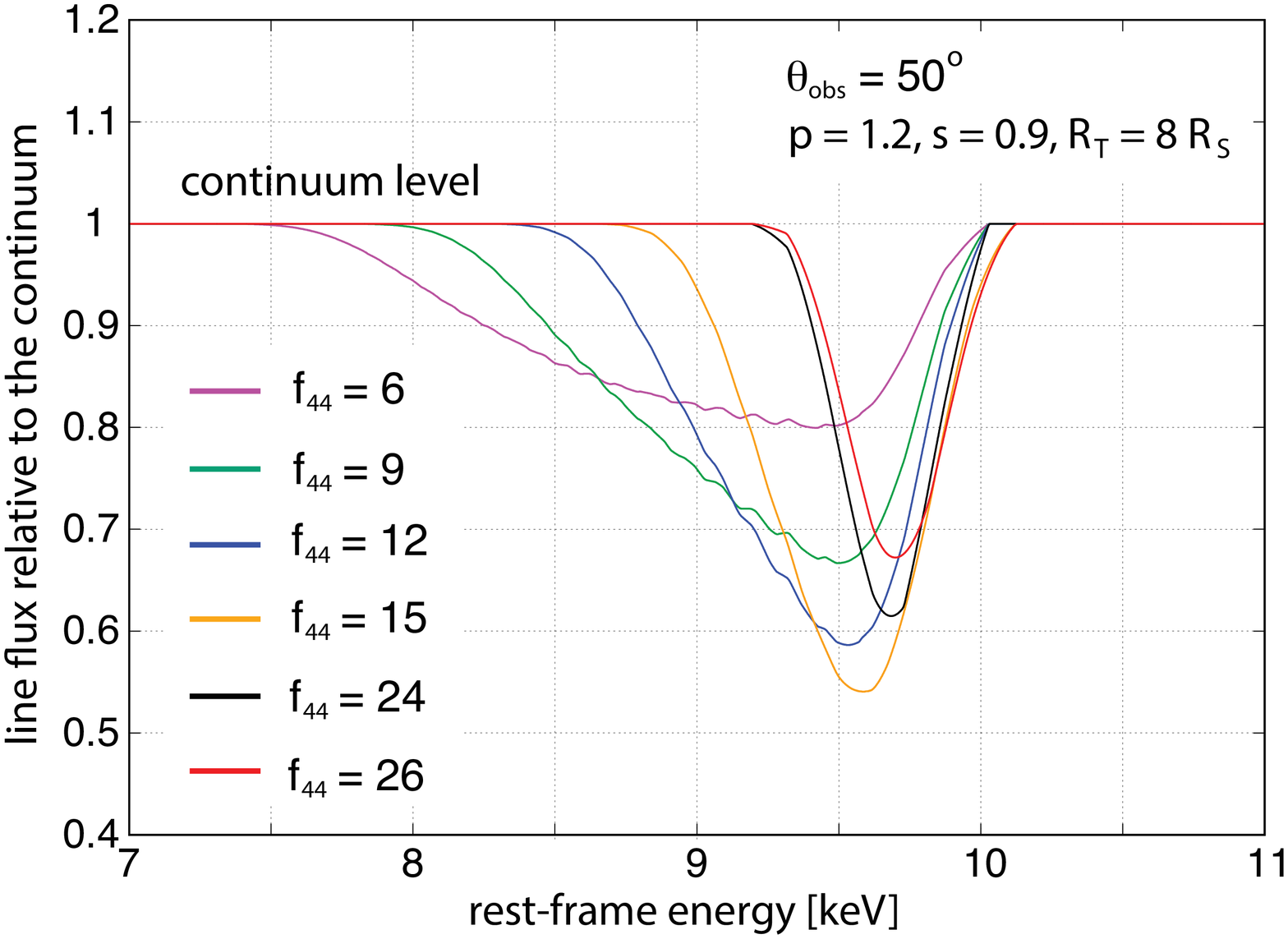}
\end{array}$
\end{center}
\caption{(a) An expected correlations between \fexxvi\ UFO velocity (red) from the MHD-driven winds 
as well as the corresponding EW (black) and the broad-band luminosity parameterized as $f_{44}$ for $s=0.9$ (solid) and $s=3$ (dashed) scalings. We have used $\theta_{\rm obs}=50\deg$ and $p=1.2$
density profile. The observed correlations are depicted by shaded green region. 
(b) Predicted template line profiles of \fexxvi\ UFOs  for different
values of $f_{44}$; $6$ (pink), $9$ (green), $12$ (blue), $15$ (orange), $24$ (black) and $26$ (red) with $\theta=50\deg$ for $R_T/R_S=8$ wind truncation and $p=1.2$
density profile.  } \label{fig:wind2}
\end{figure}

{\bf Figure~\ref{fig:wind2}a} shows  the expected correlations of the
calculated \fexxvi\ centroid velocity $v_c/c$ (red) and its
corresponding EW (black) as a function of $f_{44}$ for two cases: $s=0.9$ (solid) and
$s=3$ (dashed) for comparison.
For $s=0.9$ where wind density increases faster than the X-ray flux with $\dot{m}_a$,  
it is clearly seen that the velocity is well
correlated with $f_{44}$ while its EW is anti-correlated with $f_{44}$ \citep[i.e.
X-ray Baldwin effect;][]{IwasawaTaniguchi93} in a good agreement with data (e.g.
M17,P17,Pa18,Pi18). In this case, the UFO location (i.e. ionization front) comes closer to the BH with increasing $f_{44}$ since its radial distance scales as $r_{\rm \fexxvi} \propto
\dot{m}_a^{-1/2}$. In contrast, $s=3$ case is  ruled out in this model.
We note, for $s=0.9$, that the predicted \fexxvi\ EW exhibits a peak as $f_{44}$ varies due to the fact that the wind becomes Thomson thick. The exact peak position depends on the SED and the value 
of $s$ as the wind flux increases with $\dot m_a$. Hence, the increasing segment of 
the EW correlation with is less likely to be observed in near-Eddington sources such as
PDS 456 (shaded region in green, consistent with the data), but perhaps it is relevant in 
low/sub-Eddington Seyferts exhibiting conventional WAs and UFOs.
%

%

%
%

We show in {\bf Figure~\ref{fig:wind2}b} the
luminosity-dependence of the predicted \fexxvi\ line profiles for $6 \le f_{44} \le 26$
with $s= 0.9$. The centroid energy gradually increases with $f_{44}$, while the line
width generally decreases (thus reducing EW) {\bf qualitatively consistent with the multi-epoch {\it Suzaku} data }
\citep[e.g.][]{Matzeu16}.
This can be explained in terms of photoionization of the present
model; with increasing $\dot{m}_a$, more plasma is channeled into
the wind (i.e. $n_o^w \propto \dot{m}_a^2$ as $\eta_w \propto \dot{m}_a$) from the disk
and X-ray power increases as well (i.e. $f_{44} \propto \dot{m}_a^s$). For $\eta_w
\propto \dot m_a$, the wind density increases faster than the X-ray flux and the the
wind ionization parameter $\xi$ slowly decreases with $\dot{m}_a$.
%
%
%
%
{\bf This ionization change will inevitably prevent the wind from producing \fexxvi\ as the source brightens. Therefore, only the gas at smaller LoS radius (closer to the irradiating central X-ray source) where velocity is higher can be effectively photoionized to produce \fexxvi. The dominant heating process for such a large $\dot{m}$ is Compton scattering and electron recoil. Hence, more X-rays are scattered within the near-Compton-thick wind, depleting the photoionizing flux  in turn to suppress the efficiency of \fexxvi\ production at larger $\dot{m}$.
}
%
%
%
%
%
With further increase in $\dot{m}_a$, this effect becomes more
prominent bringing the ionization front more inwards 
where the wind is faster, while reducing the EW more. These trends are
consistent with the multi-epoch observations (M17,Pa18,Pi18).
This will eventually lead to little production of \fexxvi\ ions resulting in no
spectroscopic detection (despite the presence of high-velocity winds at all times).
%

\begin{figure}[h]
\begin{center}$
\begin{array}{cc}
\includegraphics[trim=0in 0in 0in
0in,keepaspectratio=false,width=3.5in,angle=-0,clip=false]{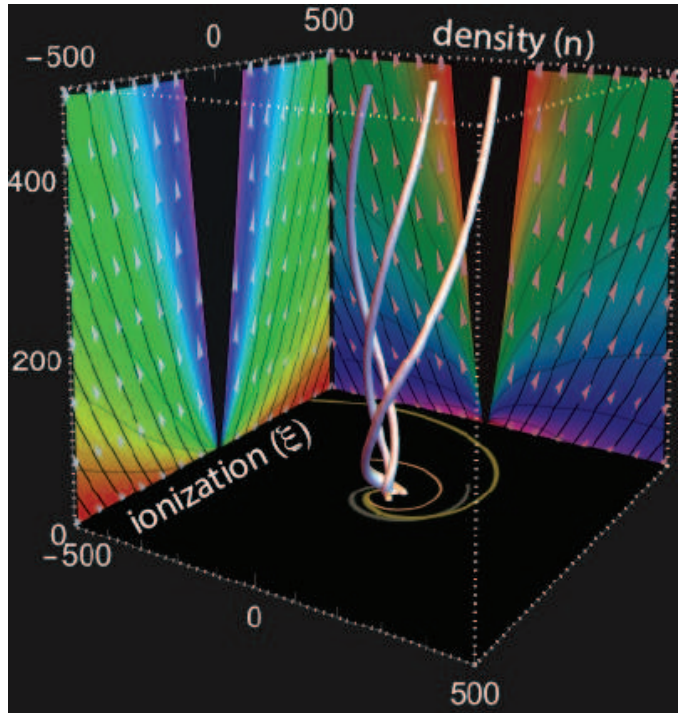}
\end{array}$
\end{center}
\caption{Three-dimensional visualization of the calculated streamlines of the wind magnetically launched from a thin disk surface.
Wind density $n(r,\theta)$ and fiducial ionization parameter $\xi(r,\theta)$ projected onto the poloidal plane (see F10 for the model description). } \label{fig:wind3}
\end{figure}

\begin{deluxetable}{l|c}
\tabletypesize{\small} \tablecaption{Inferred UFO properties with the Best-Fit MHD-Wind Model for 2013/2014 {\it XMM-Newton}/EPIC and {\it NuSTAR} Data} \tablewidth{0pt}
\tablehead{Variable & Obtained Value}
\startdata
$E_{\rm P-Cygni}$ [keV] & $6.32^{+0.13}_{-0.25}$ $^\sharp$  ($7.42$; N15) $^\diamond$ \\
Density normalization $n_{10}$ $^{\dagger}$ [$10^{10}$ cm$^{-3}$] & $9.1^{+2.8}_{-1.8}$ $^\sharp$  ($7.6^{+0.46}_{-0.42}$) $^\diamond$ \\
LoS truncated radius $R_T/R_S$ & $7.98^{+0.5}_{-1.2}$ ($7.90^{+0.3}_{-1.5}$) $^\diamond$  \\
\fexxvi\ UFO $v_{\rm c}/c$ $^\clubsuit$ & $0.35$  \\
\fexxvi\ UFO $v_{\rm max}/c$ $^\ddagger$ & $0.39$ \\
$\log (\xi_{\rm max} \textmd{[erg~cm~s$^{-1}$]})$ $^\ddagger$ & $5.5$ \\
$\log (T_{\rm max} \textmd{[K]})$ $^\ddagger$ & $5.6$ \\
$N_H^{}(\textmd{\fexxvi})$ [cm$^{-2}$] & $2.0 \times 10^{23}$ \\
$\chi^2/$dof & $382.8/430$ $^\sharp$ ~ ($501.1/429$) $^\diamond$ \\
\enddata
\vspace{0.05in}
We assume $M = 10^9 \Msun$ \citep{Reeves09}, $\theta_{\rm obs}=50\deg$ and $p=1.2$.
\\
$^\dagger$ Wind density normalization in units of $10^{10}$ cm$^{-3}$.
\\
$^\clubsuit$ Calculated value from the modeled centroid velocity.
\\
$^\ddagger$ Calculated value near the truncated radius at $r=R_T$.
\\
$^\sharp$ Treated as a free parameter. \\
$^\diamond$ A fixed value obtained from N15.
\label{tab:tab1}
\end{deluxetable}

\section{Summary \& Discussion}

We have employed the MHD accretion disk wind model introduced in our earlier works (F10, F17, F18) to account for the observed 
correlations of the \fexxvi\ UFO in \obj. 
Our model does not include explicitly the effects of radiation pressure, as do models specifically built to consider  radiatively-driven outflows \citep[e.g.][]{Higginbottom14,Hagino17,Nomura17}. These are expected to be significant in sources accreting close to their Eddington rate ($\dot m_a \simeq 1)$ like \obj.  
We have demonstrated by spectral analysis 
that the observed \fexxvi\ UFO feature of \obj\ can be successfully reproduced  within the framework the magnetically-driven disk-winds. Our MHD-wind model can also account for the observed correlations of the UFO velocity and EW with X-ray flux over multi-epoch data. The model assumes that the wind mass flux can increase faster than X-rays, a situation not unreasonable in such high mass-accretion object where radiation can be trapped and advected with the flow.  {\bf Figure~\ref{fig:wind3}} shows the
calculated streamlines along with wind density $n(r,\theta)$ and fiducial ionization
parameter $\xi(r,\theta)$ in the poloidal plane. 

%
%

The near-Eddington luminosity of \obj\ is crucial in producing an increase in the UFO velocity with $f_{44}$ since the \fexxvi\ velocity is closely related to the local escape velocity. 
An increase in this velocity implies that the ionization front responsible for the UFO moves radially inwards. This seems natural only when flows are close to Thomson-thick (a fact determined not by the X-rays but by the near-Eddington O/UV luminosity of \obj), {\bf which is conceivable as the O/UV photons are closely related to mass-accretion rate}. Of support of this notion are the observations of a near-Eddington narrow-line Seyfert AGN, IRAS~13224-3809, which exhibits similar UFO  correlations \citep{Pinto18}. 

{\bf To explore this, we have considered different density profiles and confirmed that the wind of $p=0.9$ indeed produces a smaller \fexxvi\ column (i.e. lower EW) than does the $p=1.5$ case as discussed in \S 3.2. 
}

%
Low/sub-Eddington AGNs to the contrary are not expected to have this effect, allowing the EW to increase with X-ray strength as depicted in {\bf Figure~\ref{fig:wind2}a}.
{\bf 
Overall, we note that the exact slope of the calculated correlation is sensitive to the values of $s$ and $p$ as well as the X-ray SED, and the study of such dependences will be left as a future work.    
}

Fast X-ray outflows have also been detected in BH XRBs
\citep[e.g.][]{Miller15,Miller16} that can be considered as ``scaled-down" AGN UFOs in
this framework. With timescales much shorter than those in AGNs, fast XRB winds over many binary orbits with X-ray variability may provide us with another valuable clue to
systematically understand those correlations as discussed here
for \obj.

The planned future missions, such as {\it XARM} and {\it Athena},  will  be able to
better constrain the enigmatic UFO properties with unprecedented
energy resolution, perhaps leading to  answering the ultimate
question  of its launching mechanism.

\acknowledgments We are grateful to the anonymous referee for helpful comments on the original manuscript and James Reeves for providing us with the {\it
XMM-Newton/NuSTAR} spectrum. KF thanks to Alex Sanner for his assistance on
wind visualization calculations.
This work is supported in part by NASA/ADAP (NNH15ZDA001N-ADAP) and {\it
Chandra} Cycle 17 archive proposal (17700504).

%
%
%
%
%
%

\end{document}